\documentclass[12pt,a4paper]{article}
\usepackage{amsmath}
\usepackage{amssymb}
\usepackage{epsfig}
\usepackage{subfigure}
\usepackage{mathrsfs}
\usepackage{cite}

\newcommand{\eq}[1]{Eq.~(\ref{#1})}

\newcommand{\abs}[1]{\left|#1\right|}

\newcommand{\CKM}{V}
\newcommand{\V}[1]{\CKM_{#1}^{\phantom{\ast}}}
\newcommand{\Vc}[1]{\CKM_{#1}^{\ast}}

\newcommand{\U}{U}
\newcommand{\Uu}[1]{\U_{#1}^{\phantom{\ast}}}
\newcommand{\Uuc}[1]{\U_{#1}^{\ast}}
\newcommand{\cw}{\cos\theta_W}

\newcommand{\sws}{\sin^2\theta_W}
\newcommand{\DDmix}{$D^0$--$\bar D^0$}
\newcommand{\BBdmix}{$B^0_d$--$\bar B^0_d$}
\newcommand{\BBsmix}{$B^0_s$--$\bar B^0_s$}
\newcommand{\DMD}{\Delta M_{D}}
\newcommand{\DMBs}{\Delta M_{B_s}}
\newcommand{\DMBd}{\Delta M_{B_d}}

\newcommand{\DGBdGBd}{\Delta \Gamma_{d}/\Gamma_{d}}
\newcommand{\DGBs}{\Delta \Gamma_{s}}
\newcommand{\DGBsCP}{\Delta \Gamma_{s}^{CP}}
\newcommand{\AJPsi}{A_{J/\psi K_S}}

\newcommand{\Asld}{A_{sl}^d}

\newcommand{\Asl}{\mathcal A}

\newcommand{\AJPsiP}{A_{J/\Psi\Phi}}

\newcommand{\Kpnn}{K^+\to\pi^+\nu\bar\nu}

\begin{document}

\begin{flushright}
\begin{minipage}{0.28\textwidth}
{\footnotesize{\sc IFIC/08-27\\ FTUV-08-0527\\ RM3-TH-08-10}}
\end{minipage}
\end{flushright}

\begin{center}
\begin{Large}
{\bf Small violations of unitarity, the phase in $\boldsymbol{B^0_s}$--$\boldsymbol{\bar B^0_s}$ and visible $\boldsymbol{t\to cZ}$ decays at the LHC}\\
\end{Large}
\vspace{0.5cm}
Francisco J. Botella$^{~a}$, Gustavo C. Branco$^{~b,a}$, Miguel Nebot$^{~c}$\\ \vspace{0.3cm}
{\small \emph{
$^a$ Departament de F\'{\i}sica Te\`orica and IFIC,\\ Universitat de Val\`encia-CSIC,\\ E-46100, Burjassot, Spain\\
$^b$ Centro de F\'{\i}sica Te\'orica de Part\'{\i}culas (CFTP),\\ Instituto Superior T\'ecnico,\\ P-1049-001, Lisboa, Portugal\\
$^c$ Istituto Nazionale di Fisica Nucleare (INFN), Sezione di Roma Tre,\\
\& Dipartimento di Fisica ``\emph{Edoardo Amaldi}'', Universit\`a degli Studi Roma Tre, I-00146, Roma, Italy.
}}
\end{center}

\begin{abstract}
We show that it is possible to accommodate the observed size of the phase in \BBsmix\, mixing in the framework of a model with violation of $3\times 3$ unitarity. This violation is associated to the presence of a new $Q=2/3$ isosinglet quark $T$, which mixes both with $t$ and $c$ and has a mass not exceeding 500 GeV. The crucial point is the fact that this framework allows for $\chi\equiv\arg(-\V{ts}\V{cb}\Vc{tb}\Vc{cs})$ of order $\lambda$, to be contrasted with the situation in the Standard Model, where $\chi$ is constrained to be of order $\lambda^2$. We point out that this scenario implies rare top decays $t\to cZ$ at a rate observable at the LHC and $|\V{tb}|$ significantly different from unity. In this framework, one may also account for the observed size of \DDmix\, mixing without having to invoke long distance contributions. It is also shown that in the present scenario, the observed size of \DDmix\ mixing constrains $\chi^\prime\equiv\arg(-\V{cd}\V{us}\Vc{cs}\Vc{ud})$ to be of order $\lambda^4$, which is significantly smaller than what is allowed in generic models with violations of $3\times 3$ unitarity.
\end{abstract}

\newpage

\section{Introduction}\label{SEC:01}

The Standard Model (SM) describes flavour mixing and CP violation through a $3\times 3$ unitary CKM matrix, characterized by four independent parameters. At present, the SM and its built-in CKM mechanism \cite{Cabibbo:1963yzKobayashi:1973fv} for mixing and CP violation are in good agreement with most of the experimental data \cite{ckmfitterutfit}, which constrain the size of New Physics \cite{Ligeti:2004akBotella:2005fcVelascoSevilla:2006dyLigeti:2006pmBall:2006xxBona:2006sa,Botella:2006va}.
This is a remarkable achievement, in view of the enormous amount of data which has to be described by the four independent CKM parameters.    Recently, some experimental evidence hint at potential deviations from the SM predictions: the flavour-tagged determination of mixing induced CP violation in $B_s\to J/\Psi \Phi$ decays \cite{Aba:2008fj,Aaltonen:2007he} and eventually the measurement of \DDmix\ mixing \cite{Aubert:2007wfStaric:2007dtAbe:2007rd}, if long distance contributions to the mixing are not dominant. There are also hints related to $b\to s$ penguin transitions \cite{Buchalla:2008jpFleischer:2008uj}.

In this letter, we address the question of whether it is possible to find a New Physics (NP) explanation for these recent experimental results. The assumption of $3\times 3$ unitarity of the CKM matrix, is one of the crucial ingredients in most of the tests of the flavour sector of the SM. Therefore, we find it reasonable to analyse the above question in the framework of models where there are small violations of $3\times 3$ unitarity of the CKM matrix. It is clear that any plausible extension of the SM where deviations of unitarity occur, should provide a natural explanation for the smallness of these violations. This is crucial, since deviations of $3\times 3$ unitarity of the CKM matrix automatically lead to flavour-changing neutral currents (FCNC) couplings which are known from experiment to be severely suppressed. The simplest models with naturally small violations of $3\times 3$ unitarity are those with vector-like isosinglet quarks \cite{vectorlike,Branco:1992wr,Lavoura:1992qd,vectorlike2,AguilarSaavedra:2002kr,AguilarSaavedra:2004mt}. Since both the left and the right components of these quarks are $SU(2)$ singlets, their mass scale $M$ can be substantially larger than the mass scale of standard quarks. In this class of models, there are violations of $3\times 3$ unitarity which are naturally suppressed by the ratio $m^2/M^2$. These violations of unitarity in quark mixing are entirely analogous to violations of unitarity which are present in the leptonic mixing matrix, when the seesaw mechanism \cite{seesaw} is used to explain the smallness of neutrino masses. We will show that within the framework of $Q=2/3$ vector like quarks, one may account for both the size of \DDmix\ mixing, as well as the value of the phase in \BBsmix\ mixing. The crucial point is that in the above framework, some of the exact SM unitarity relations \cite{Botella:2002fr}, connecting moduli and rephasing invariant phases are significantly modified, in the presence of small unitarity violations. In particular, we show that the rephasing invariant phase $\chi\equiv\arg(-\V{ts}\V{cb}\Vc{tb}\Vc{cs})$ can be much larger than in the SM, giving rise to a value compatible with the mixing induced CP asymmetry in $B_s\to J/\Psi \Phi$. We point out that a clear-cut prediction of this scenario are top decays $t\to cZ$ at a rate observable at the LHC.
This paper is organized as follows. In the next section we briefly describe the framework and set our notation. In section \ref{SEC:03}, we point out the main physical implications of small violations of $3\times 3$ unitarity, in the context of the present framework. Specific examples are given in section \ref{SEC:04}, with detailed analyses of their physical consequences. Finally, our conclusions are contained in section \ref{SEC:CONC}.

\section{Notation and framework}\label{SEC:02}

For definiteness, let us consider an extension of the SM where only up-type isosinglet quarks are added to the spectrum. In this case, the $3\times 3$ quark mixing matrix connecting standard quarks is a submatrix of a larger unitary matrix $\U$. Without loss of generality, one can choose a weak basis where the down quark mass matrix is diagonal, real. In this basis, $\U$ is just the $4\times 4$ unitary matrix which enters the diagonalization of the up quark mass matrix. With no loss of generality, one can also use the freedom to rephase quark fields, to choose the phases of $\U$ in the following way:
\begin{equation}
 \arg \U=\begin{pmatrix}0 & \chi^\prime& -\gamma & \cdots\\ \pi & 0& 0& \cdots\\ -\beta & \pi+\chi & 0& \cdots\\ \vdots &\vdots&\vdots&\ddots \end{pmatrix}~,\label{EQ:CKMphases01}
\end{equation}
where the four rephasing invariant phases are\footnote{Some authors use $\beta_s\equiv\chi$, $\phi_1\equiv\beta$ and $\phi_3\equiv\gamma$; $\chi^\prime$ is usually neglected.} \cite{Aleksan:1994if,Branco:1999fs}:
\begin{eqnarray}
 \beta\equiv \arg(-\V{cd}\Vc{cb}\Vc{td}\V{tb})&;\quad& \gamma\equiv \arg(-\V{ud}\Vc{ub}\Vc{cd}\V{cb});\nonumber\\
 \chi\equiv \arg(-\V{ts}\Vc{tb}\Vc{cs}\V{cb})&;\quad& \chi^\prime\equiv \arg(-\V{cd}\Vc{cs}\Vc{ud}\V{us}).\label{EQ:CKMphases02}
\end{eqnarray}
It should be emphasized that independently of the dimensions of $\U$, only the four rephasing invariant phases in \eq{EQ:CKMphases02} enter its $3\times 3$ sector connecting standard quarks. In the three generations SM, these four rephasing invariant phases and the nine moduli of $\CKM$ are related by various exact relations \cite{Botella:2002fr} which provide a test of the SM. It can be readily verified that in the context of the SM, the phases $\chi$ and $\chi^\prime$ are small, of order $\lambda^2$ and $\lambda^4$, respectively, with $\lambda\simeq 0.2$. It has been pointed out that in the framework of models with up-type isosinglet quarks \cite{AguilarSaavedra:2004mt}, one can obtain larger values of $\chi$, of order $\lambda$.

 As above mentioned, we assume that there is only one up-type isosinglet quark, which we denote T. In the mass eigenstate basis the charged and neutral current interactions can be written:
\begin{eqnarray}
 \mathscr L_W &=&-\frac{g}{\sqrt 2}\bar {\mathbf u}_L \gamma^\mu \CKM {\mathbf d}_L W_\mu^\dagger+\text{H.C.}~,\nonumber\\
 \mathscr L_Z &=&-\frac{g}{2\cw}\left[\bar {\mathbf u}_L \gamma^\mu (\CKM\CKM^\dagger){\mathbf u}_L-\bar {\mathbf d}_L \gamma^\mu {\mathbf d}_L-2\sws J^\mu_{em}\right]Z_\mu~,\label{EQ:LWLZ}
\end{eqnarray}
where $\mathbf u=(u,c,t,T)$, $\mathbf d=(d,s,b)$, while $\CKM$ is a $4\times 3$ submatrix of the $4\times 4$ unitary matrix $\U$ which enters the diagonalization of the up-type quark mass matrix:
\begin{equation}
 \CKM=\begin{pmatrix}\V{ud}& \V{us}& \V{ub}\\\V{cd}& \V{cs}& \V{cb}\\ \V{td}& \V{ts}& \V{tb}\\ \V{Td}& \V{Ts}& \V{Tb}\end{pmatrix}~. \label{EQ:CKMmatrix}
\end{equation}
It is clear from Eqs. (\ref{EQ:LWLZ},~\ref{EQ:CKMmatrix}), that $\CKM\CKM^\dagger\neq \mathbf 1$, which leads to FCNC in the up-quark sector. The salient feature of this class of models with isosinglet quarks is that there are naturally small violations of unitarity. It is clear from \eq{EQ:CKMmatrix} that the columns of $\CKM$ are orthogonal, while its rows are not. 

\section{Physical implications of small violations of $\boldsymbol{3\times 3}$ unitarity}\label{SEC:03}

We analyse next the most salient consequences of having small violations of unitarity. Although our analysis is done within the framework of one isosinglet quark $T$, a good part of our results hold in a much larger class of extensions of the SM. The crucial ingredient is the presence of small violations of unitarity, independently of their origin.

\subsection{Obtaining a large $\boldsymbol\chi$}\label{sSEC:0301}

From orthogonality of the second and third column of $\CKM$, one obtains \cite{AguilarSaavedra:2004mt}:
\begin{equation}
 \sin\chi=\frac{\abs{\V{ub}}\abs{\V{us}}}{\abs{\V{cb}}\abs{\V{cs}}}\sin(\gamma-\chi+\chi^\prime)+\frac{\abs{\V{Tb}}\abs{\V{Ts}}}{\abs{\V{cb}}\abs{\V{cs}}}\sin(\sigma-\chi)~,\label{EQ:32Corth}
\end{equation}
where $\sigma$ is a rephasing invariant phase, $\sigma\equiv \arg(\V{Ts}\V{cb}\Vc{Tb}\Vc{cs})$. It is clear that $\chi$ can be of order $\lambda$ if one has, for example, $\V{Tb}\approx \mathcal O(\lambda)$, $\V{Ts}\approx \mathcal O(\lambda^2)$, $\sigma\approx \mathcal O(1)$. In the SM one has, of course, $\sin\chi=\mathcal O(\lambda^2)$, since only the first term in \eq{EQ:32Corth} is present. The crucial conclusion from \eq{EQ:32Corth} is that in order to obtain $\chi$ of order $\lambda$, the coupling of the $T$ quark to the bottom quark has to be of the size of Cabibbo mixing, while the coupling $T$ to the strange quark has to be of the same order as $\V{ts}$. We shall see through the examples that these values of $\V{Tb}$, $\V{Ts}$ can accommodate the present experimental constraints. It is further required that the phase of the rephasing invariant $\sigma$ is of order one, which in the phase convention of \eq{EQ:CKMphases01} implies that the phase of the bilinear $\V{Ts}\Vc{Tb}$ is large. So far, we have not exploited the connection between the size of $\chi$ and violations of unitarity arising from non-orthogonality among the rows of $\CKM$. In order to see this connection, let us consider orthogonality of the second and third rows of the $4\times 4$ unitary matrix $\U$. One obtains:
\begin{equation}
 \sin\chi=\frac{\abs{\V{cd}}\abs{\V{td}}}{\abs{\V{cs}}\abs{\V{ts}}}\sin\beta+\frac{\abs{\Uu{24}}\abs{\Uu{34}}}{\abs{\V{cs}}\abs{\V{ts}}}\sin\delta~,\label{EQ:32Rorth}
\end{equation}
where $\delta$ is the rephasing invariant phase $\delta=\arg(\Vc{tb}\Uuc{24}\V{cb}\Uu{34})$. It is clear that in order to have $\chi=\mathcal O(\lambda)$, having $\abs{\Uu{24}}\abs{\Uu{34}}$ of order $\lambda^3$ is required, for example:
\begin{equation}
 \abs{\Uu{24}}\approx \mathcal O(\lambda^2),\qquad \abs{\Uu{34}}\approx \mathcal O(\lambda),\qquad \sin\delta\approx \mathcal O(1)~.\label{EQ:LambdaChi}
\end{equation}
This has important implications for rare top decays which we analyse in the next subsection.

\subsection{Rare top decays}\label{sSEC:0302}

From \eq{EQ:LWLZ} it follows that the FCNC couplings of the type $\bar c_L \gamma^\mu t_L\, Z_\mu$ are proportional to $|\Uu{24}\Uu{34}|$, which measures deviations from orthogonality of the second and third rows of $\CKM$. On the other hand, as pointed out above, \eq{EQ:32Rorth} tells us that having $\chi$ of order $\lambda$ requires $|\Uu{24}\Uu{34}|$ of order $\lambda^3$. This in turn implies a significant $\bar c_L \gamma^\mu t_L Z_\mu$ coupling which leads to rare top decays $t\to cZ$, at rates such that they can be observed at the LHC. Therefore, in the framework of this model, one finds the following important correlation: a large value of $\chi$ necessarily implies rare top decays of the type $t\to cZ$, visible at the LHC. This is a distinctive feature of this framework.

\subsection{Contributions to $\boldsymbol{D^0}$-- $\boldsymbol{\bar D^0}$ mixing}\label{sSEC:0303}

As previously mentioned, in the present framework there are FCNC in the up-quark sector, contributing to $Z$ couplings to $\bar c_L \gamma^\mu u_L$ at tree level \cite{Branco:1995us}. In order for these couplings to be able to account for the observed size of \DDmix, without having to invoke long-distance contributions to the mixing, the size of $|\Uu{14}\Uu{24}|$ has to be of order $\lambda^5$ \cite{Golowich:2007ka}. On the other hand, we have seen that in order to obtain $\chi$ of order $\lambda$ in this framework $|\Uu{24}|$ is required to be of order $\lambda^2$ which leads to $|\Uu{14}|=\mathcal O(\lambda^3)$. We shall see in the next subsection that this value of $|\Uu{14}|$ is considerably smaller than the upper bounds derived from normalization of the first row of $\U$. Therefore, in the framework of this model, one may find an explanation for the size of \DDmix\ without violating the bound on $|\Uu{14}|$. Obviously, since we only have an upper bound on $|\Uu{14}|$, even in a scenario with $\chi=\mathcal O(\lambda)$, one may also have a negligible New Physics contribution to \DDmix\ mixing by choosing $|\Uu{14}|$ sufficiently small. In that case, long distance contributions would be responsible for the observed size of \DDmix\ mixing.

\subsection{Deviations of unitarity in the first row of $\boldsymbol{\CKM}$ and the size of $\boldsymbol{\chi^\prime}$.}\label{sSEC:0304}

 At present, the most precisely measured elements of $\CKM$ are $|\V{ud}|$, $|\V{us}|$ \cite{Towner:2007npEronen:2007qc,Antonelli:2008jg}:
\begin{eqnarray*}
 |\V{ud}|&=& 0.97408\pm 0.00026~,\\
 |\V{us}|&=& 0.2253\pm 0.0009~.
\end{eqnarray*}
This leads to the following upper bound
\begin{equation*}
 1-\left(|\V{ud}|^2+|\V{us}|^2+|\V{ub}|^2\right)=|\Uu{14}|^2\lesssim (0.02)^2~.
\end{equation*}
It is clear that $|\Uu{14}|$ is restricted by the precise measurements of the first row of $\CKM$. However, as mentioned in the previous subsection, this bound on $|\Uu{14}|$ is less restrictive than the one derived from the size of \DDmix\ mixing in the framework of $\chi=\mathcal O(\lambda)$, which requires $|\Uu{24}|=\mathcal O(\lambda^2)$. Next, we show how the size of $\chi^\prime\equiv\arg(-\V{cd}\V{us}\Vc{cs}\Vc{ud})$ is constrained, in the present framework, by the size of \DDmix\ mixing. From orthogonality of the first two rows of $\U$, one readily obtains the exact relation:
\begin{equation}
 \sin\chi^\prime=\frac{|\V{ub}\V{cb}|}{|\V{us}\V{ub}|}\sin\gamma+\frac{|\Uu{14}\Uu{24}|}{|\V{us}\V{ub}|}\sin\rho~,\label{EQ:U12orth}
\end{equation}
where $\rho\equiv\arg(\V{cd}\Uu{14}\Vc{ud}\Uuc{24})$. This is a generalization of the relation for $\sin\chi^\prime$ in the context of the SM. The first term in the right-hand side of \eq{EQ:U12orth} is, of course, of order $\lambda^4$. The interesting point is that in models with an isosinglet $T$ quark, $|\Uu{14}\Uu{24}|$ is restricted by the size of \DDmix\ mixing to be at most $\mathcal O(\lambda^5)$ and thus the second term of \eq{EQ:U12orth} is also of order $\lambda^4$ and as a result $\sin\chi^\prime\sim\mathcal O(\lambda^4)$. In generic models with violations of $3\times 3$ unitarity $\sin\chi^\prime$ can be significantly larger.

\section{Examples}\label{SEC:04}
In this section we will present some examples of mixing matrices $\CKM$ which have been obtained by imposing the following requirements\footnote{A summary of the experimental values employed is given in appendix \ref{AP:signCHI}.}:
\begin{itemize}
 \item sizable mixing induced, time dependent, CP-violating asymmetry in $B_s^0\to J/\Psi \Phi$ for the CP-even part of the final state, $\AJPsiP\equiv \sin 2\chi_{eff}$, where $2\chi_{eff}=-\arg{M_{12}^{B_s}}$. For this observable, already accessible at Tevatron, we consider in particular the last D$\emptyset$ result $0.540\pm 0.225$ \cite{Aba:2008fj}.
 \item the contribution to $x_D\equiv \DMD/\Gamma_D$ in \DDmix\ mixing from tree level FCNC mostly accounts for the observed value \cite{Aubert:2007wfStaric:2007dtAbe:2007rd}. Apart from the previous contribution -- 'short distance' originated --, there might be important long distance ones. Therefore, examples giving significantly smaller contributions to $x_D$ are also shown, illustrating, as already mentioned, that large $x_D$ are not compulsory in the scenario under study.
 \item agreement with purely tree level observables constraining $\CKM$, namely, moduli in the first two rows and the physical phase $\gamma$.
 \item agreement with the following observables potentially sensitive to New Physics \cite{Botella:2006va,AguilarSaavedra:2002kr,AguilarSaavedra:2004mt}:
  \begin{itemize}
   \item mixing induced, time dependent, CP-violating asymmetry in $B_d^0\to J/\Psi K_S$ vs. $B_d^0\to \bar B_d^0\to J/\Psi K_S$, $\AJPsi$.
   \item mass differences $\DMBd$, $\DMBs$, of the eigenstates of the effective Hamiltonians controlling \BBdmix\ and \BBsmix\ mixings. 
   \item width differences $\DGBdGBd$, $\DGBs$, $\DGBsCP$ of the eigenstates of the mentioned effective Hamiltonians, related to $\text{Re}\left(\Gamma^q_{12}/M^q_{12}\right)$, $q=d,s$.
   \item charge/semileptonic asymmetries $\Asl$, $\Asld$, controlled by $\text{Im}\left(\Gamma^q_{12}/M^q_{12}\right)$, $q=d,s$.
   \item neutral kaon CP violating parameters $\epsilon_K$ and $\epsilon^\prime/\epsilon_K$ \cite{eprime}.
   \item branching ratios of representative rare K and B decays such as $K^+\to\pi^+\nu\bar\nu$, $(K_L\to\mu\bar\mu)_{SD}$ and $B\to X_s\ell^+\ell^-$ \cite{Antonelli:2008jg,Marciano:1996wyBuchalla:1998ba,klmumu,Huber:2005igHuber:2007vv}.
   \item electroweak oblique parameter $T$, which encodes violation of weak isospin; the $S$ and $U$ parameters play no relevant r\^ole here \cite{Lavoura:1992qd}. Note that in the region we are interested in, other precision electroweak parameters like $R_b$ give similar constraints to those obtained from $T$ \cite{Alwall:2006bxPicek:2008dd}.
  \end{itemize}
 \item beside those experimentally based constraints, agreement with the available results is required for every parameter entering the calculation of the observables, like QCD corrections, lattice-QCD computed bag factors, etc.
\end{itemize}

\subsection{Example 1, $\boldsymbol{m_T=300}$ GeV and large $\boldsymbol{x_D}$}\label{sSEC:EX01}
For the first example we have a rather light value for the mass of the additional up-type quark, $m_T=300$ GeV, the following moduli of $\U\supset\CKM$:
\begin{equation}
 |\U|=\left(\begin{matrix}0.974186& 0.225642& 0.003984\ \\ 0.225559& 0.972463& 0.041676\ \\ 0.009002& 0.047563& 0.948582\ \\ 0.001666& 0.033749& 0.313759\ \end{matrix}\right|\left.\begin{matrix} 0.005530\\ 0.041252\\ 0.312809\\ 0.948904\end{matrix}\right)~,\label{eq:EX01mod}
\end{equation}
and the phases (presented here and in the following in an easily readable phase convention):
\begin{equation}
 \arg{\U}=\left(\begin{matrix}0& 0.000530& -1.055339\ \\ 
 \pi& 0& 0\ \\ -0.472544& \pi-0.208060& 0\ \\ 1.795665& -1.266410& 0\ \end{matrix}\right|\left.\begin{matrix}1.071901 \\ 0.947622\\ 0\\ \pi+0.004752 \end{matrix}\right)~.\label{eq:EX01arg}
\end{equation}
The set of relevant observables that follow are displayed in table \ref{TAB:EX01}.
\begin{table}[h]
\begin{center}
\begin{tabular}{|c|c||c|c|}
\hline
  Observable & Value & Observable & Value \\ \hline\hline
\small{$\gamma$}& \small{$60.5^\circ$}&\small{$\chi$} &\small{$-11.9^\circ$}\\\hline
\small{$\DMBd$} & \small{0.507 ps$^{-1}$} & \small{$\DMBs$}& \small{17.77 ps$^{-1}$}\\\hline
\small{$\AJPsi$} & \small{0.692} &  \small{$\AJPsiP$}& \small{0.288}\\\hline
\small{$\epsilon_K$} & \small{$2.232\times 10^{-3}$}& \small{$\epsilon^\prime/\epsilon_K$} & \small{$1.63\times 10^{-3}$} \\\hline
\small{$x_D$} & \small{0.0085} & \small{$\Delta T$} &  \small{0.16} \\\hline
\small{Br$(\Kpnn)$} &  \small{$1.3\times 10^{-10}$} &  \small{Br$(K_L\to\mu\bar\mu)_{SD}$} & \small{$1.86\times 10^{-9}$}\\\hline
\small{Br($t\to cZ$)} & \small{$1.4\times 10^{-4}$} & \small{Br($t\to uZ$)} & \small{$2.5\times 10^{-6}$} \\\hline
\small{Br$(B\to X_s e^+e^-)$} & \small{$1.63\times 10^{-6}$} & \small{Br$(B\to X_s \mu^+\mu^-)$} & \small{$1.58\times 10^{-6}$}\\\hline
\small{$\DGBdGBd$} & \small{$0.0042$} \\\hline
\small{$\DGBs$}& \small{0.098 ps$^{-1}$}& \small{$\DGBsCP$}& \small{0.094 ps$^{-1}$} \\\hline
\small{$\Asld$} & \small{$-0.0010$} &\small{$\Asl$} & \small{$-0.0006$}\\\hline
\end{tabular}
\caption{Observables for example 1.}\label{TAB:EX01}
\end{center}
\end{table}

\subsection{Example 2, $\boldsymbol{m_T=300}$ GeV and $\boldsymbol{x_D}$ not large}\label{sSEC:EX02}
The next example also corresponds to a mass $m_T=300$ GeV but short distance contributions alone do not  produce a value of $x_D$ close to the experimental result. The moduli of $\U\supset\CKM$ are now:
\begin{equation}
 |\U|=\left(\begin{matrix}0.974195& 0.225663& 0.004137\ \\ 0.225482& 0.972938& 0.041548\ \\ 0.009721& 0.042034& 0.945531\ \\ 0.002889& 0.026471& 0.322842\ \end{matrix}\right|\left.\begin{matrix} 0.002015\\ 0.028688\\ 0.322660\\ 0.946078\end{matrix}\right)~,\label{eq:EX02mod}
\end{equation}
while the corresponding phases:
\begin{equation}
 \arg{\U}=\left(\begin{matrix}0& 0.000569& -1.204546\ \\ 
 \pi& 0& 0\ \\ -0.536152& \pi-0.189787& 0\ \\ 1.545539& -1.774240& 0\ \end{matrix}\right|\left.\begin{matrix}1.928448 \\ 1.267846\\ 0\\ \pi+0.003725 \end{matrix}\right)~.\label{eq:EX02arg}
\end{equation}
They yield the observables in table \ref{TAB:EX02}.
\begin{table}[h]
\begin{center}
\begin{tabular}{|c|c||c|c|}
\hline
 Observable & Value & Observable & Value \\ \hline\hline
\small{$\gamma$}& \small{$69.0^\circ$}&\small{$\chi$} &\small{$-10.9^\circ$}\\\hline
\small{$\DMBd$} & \small{0.507 ps$^{-1}$} & \small{$\DMBs$}& \small{17.77 ps$^{-1}$}\\\hline
\small{$\AJPsi$} & \small{0.686} &  \small{$\AJPsiP$}& \small{0.250}\\\hline
\small{$\epsilon_K$} & \small{$2.232\times 10^{-3}$}& \small{$\epsilon^\prime/\epsilon_K$} & \small{$1.66\times 10^{-3}$} \\\hline
\small{$x_D$} & \small{0.0005} & \small{$\Delta T$} &  \small{0.17} \\\hline
\small{Br$(\Kpnn)$} &  \small{$1.2\times 10^{-10}$} &  \small{Br$(K_L\to\mu\bar\mu)_{SD}$} & \small{$1.99\times 10^{-9}$}\\\hline
\small{Br($t\to cZ$)} & \small{$0.72\times 10^{-4}$} & \small{Br($t\to uZ$)} & \small{$3.5\times 10^{-7}$} \\\hline
\small{Br$(B\to X_s e^+e^-)$} & \small{$1.92\times 10^{-6}$} & \small{Br$(B\to X_s \mu^+\mu^-)$} & \small{$1.86\times 10^{-6}$}\\\hline
\small{$\DGBdGBd$} & \small{$0.0042$}  \\\hline
\small{$\DGBs$}& \small{0.088 ps$^{-1}$}& \small{$\DGBsCP$}& \small{0.085 ps$^{-1}$} \\\hline
\small{$\Asld$} & \small{$-0.0013$} &\small{$\Asl$} & \small{$-0.0006$}\\\hline
\end{tabular}
\caption{Observables for example 2.}\label{TAB:EX02}
\end{center}
\end{table}

Beside the previous examples with $m_T=300$ GeV, examples with the additional up-type quark being heavier are also interesting; we now present a couple of them for $m_T=450$ GeV.

\subsection{Example 3, $\boldsymbol{m_T=450}$ GeV and large $\boldsymbol{x_D}$}\label{sSEC:EX03}
A first example producing large $x_D$ with $m_T=450$ GeV has the following mixing matrix moduli:
\begin{equation}
 |\U|=\left(\begin{matrix}0.974179& 0.225657& 0.004031\ \\ 0.225619& 0.972525& 0.041766\ \\ 0.008330& 0.047219& 0.966377\ \\ 0.001136& 0.032304& 0.253683\ \end{matrix}\right|\left.\begin{matrix} 0.006073\\ 0.039324\\ 0.252620\\ 0.966747\end{matrix}\right)~,\label{eq:EX03mod}
\end{equation}
 and phases:
\begin{equation}
 \arg{\U}=\left(\begin{matrix}0& 0.000570& -0.957178\ \\ 
 \pi& 0& 0\ \\ -0.447359& \pi-0.140403& 0\ \\ 1.908192& -1.055192& 0\ \end{matrix}\right|\left.\begin{matrix}0.868831 \\ 0.816488\\ 0\\ \pi+0.004977 \end{matrix}\right)~.\label{eq:EX03arg}
\end{equation}

\begin{table}[h]
\begin{center}
\begin{tabular}{|c|c||c|c|}
\hline
Observable & Value & Observable & Value \\ \hline\hline
\small{$\gamma$}& \small{$54.8^\circ$}&\small{$\chi$} &\small{$-8.0^\circ$}\\\hline
\small{$\DMBd$} & \small{0.507 ps$^{-1}$} & \small{$\DMBs$}& \small{17.77 ps$^{-1}$}\\\hline
\small{$\AJPsi$} & \small{0.693} &  \small{$\AJPsiP$}& \small{0.317}\\\hline
\small{$\epsilon_K$} & \small{$2.232\times 10^{-3}$}& \small{$\epsilon^\prime/\epsilon_K$} & \small{$1.63\times 10^{-3}$} \\\hline
\small{$x_D$} & \small{0.0092} & \small{$\Delta T$} &  \small{0.20} \\\hline
\small{Br$(\Kpnn)$} &  \small{$1.0\times 10^{-10}$} &  \small{Br$(K_L\to\mu\bar\mu)_{SD}$} & \small{$1.87\times 10^{-9}$}\\\hline
\small{Br($t\to cZ$)} & \small{$0.80\times 10^{-4}$} & \small{Br($t\to uZ$)} & \small{$1.88\times 10^{-6}$} \\\hline
\small{Br$(B\to X_s e^+e^-)$} & \small{$1.60\times 10^{-6}$} & \small{Br$(B\to X_s \mu^+\mu^-)$} & \small{$1.55\times 10^{-6}$}\\\hline
\small{$\DGBdGBd$} & \small{$0.0041$}\\\hline
\small{$\DGBs$}& \small{0.110 ps$^{-1}$}& \small{$\DGBsCP$}& \small{0.104 ps$^{-1}$} \\\hline
\small{$\Asld$} & \small{$-0.0010$} &\small{$\Asl$} & \small{$-0.0007$}\\\hline
\end{tabular}
\caption{Observables for example 3.}\label{TAB:EX03}
\end{center}
\end{table}

\subsection{Example 4, $\boldsymbol{m_T=450}$ GeV and $\boldsymbol{x_D}$ not large}\label{sSEC:EX04}
For completeness let us display the last example, also with $m_T=450$ GeV, but yielding much smaller $x_D$. The mixing matrix moduli are:
\begin{equation}
 |\U|=\left(\begin{matrix}0.974192& 0.225675& 0.004015\ \\ 0.225535& 0.972984& 0.041642\ \\ 0.009033& 0.044207& 0.961556\ \\ 0.001741& 0.020444& 0.271403\ \end{matrix}\right|\left.\begin{matrix} 0.002260\\ 0.026487\\ 0.270876\\ 0.962247\end{matrix}\right)~,\label{eq:EX04mod}
\end{equation}
 and the phases:
\begin{equation}
 \arg{\U}=\left(\begin{matrix}0& 0.000622& -1.092316\ \\ 
 \pi& 0& 0\ \\ -0.467721& \pi-0.108029& 0\ \\ 1.920727& -1.329417& 0\ \end{matrix}\right|\left.\begin{matrix}1.085654 \\ 0.885746\\ 0\\ \pi+0.003299 \end{matrix}\right)~.\label{eq:EX04arg}
\end{equation}

\begin{table}[h]
\begin{center}
\begin{tabular}{|c|c||c|c|}
\hline
Observable & Value & Observable & Value \\ \hline\hline
\small{$\gamma$}& \small{$62.6^\circ$}&\small{$\chi$} &\small{$-6.2^\circ$}\\\hline
\small{$\DMBd$} & \small{0.507 ps$^{-1}$} & \small{$\DMBs$}& \small{17.77 ps$^{-1}$}\\\hline
\small{$\AJPsi$} & \small{0.688} &  \small{$\AJPsiP$}& \small{0.265}\\\hline
\small{$\epsilon_K$} & \small{$2.232\times 10^{-3}$}& \small{$\epsilon^\prime/\epsilon_K$} & \small{$1.66\times 10^{-3}$} \\\hline
\small{$x_D$} & \small{0.0006} & \small{$\Delta T$} &  \small{0.23} \\\hline
\small{Br$(\Kpnn)$} &  \small{$1.0\times 10^{-10}$} &  \small{Br$(K_L\to\mu\bar\mu)_{SD}$} & \small{$2.10\times 10^{-9}$}\\\hline
\small{Br($t\to cZ$)} & \small{$0.42\times 10^{-5}$} & \small{Br($t\to uZ$)} & \small{$3.0\times 10^{-7}$} \\\hline
\small{Br$(B\to X_s e^+e^-)$} & \small{$1.75\times 10^{-6}$} & \small{Br$(B\to X_s \mu^+\mu^-)$} & \small{$1.70\times 10^{-6}$}\\\hline
\small{$\DGBdGBd$} & \small{$0.0043$}\\\hline
\small{$\DGBs$}& \small{0.098 ps$^{-1}$}& \small{$\DGBsCP$}& \small{0.094 ps$^{-1}$} \\\hline
\small{$\Asld$} & \small{$-0.0012$} &\small{$\Asl$} & \small{$-0.0006$}\\\hline
\end{tabular}
\caption{Observables for example 4.}\label{TAB:EX04}
\end{center}
\end{table}

\subsection{Comments}\label{sSEC:04:Results}

At this stage the following comments are in order.\newline

{\bf $\boldsymbol{B^0_s}$--$\boldsymbol{\bar B^0_s}$ mixing phase}\\

The examples presented in the previous section provide values of the CP violating asymmetry $\AJPsiP$ in the range $[0.25;0.32]$, significantly larger than the SM expectation $0.04$. These values are in agreement with the D$\emptyset$ result. However the model does not allow for much larger values of $\AJPsiP$, even if one considers values $m_T$ that are larger, but not excessively larger in order to avoid difficulties with non-decoupling contributions to flavour changing rare decays.
Notice also that $\AJPsiP$ \emph{is not} $\sin 2\chi$ \emph{but} $\sin 2\chi_{eff}$ (see appendix \ref{AP:signCHI}).\newline

\newpage
{\bf $\boldsymbol{D^0}$--$\boldsymbol{\bar D^0}$ mixing}\\

Concerning \DDmix\ mixing we have presented examples that could account for $x_D$ just through the short-distance contribution present in this scenario: the tree level $Z$-mediated one. This is not, as stressed above, compulsory: examples with short-distance contributions not accounting for $x_D$ are as well presented. The crucial test to disentangle the origin of \DDmix\ mixing -- short or large distance --, could come from CP violation, and the present model certainly produces new CP-violating phases.\newline

{\bf Observable $\boldsymbol{t\to c Z}$ decays at the LHC and $\boldsymbol{|\V{tb}|}\neq 1$}\\

The branching ratio of $t\to c Z$ decays has, in the above examples, values $10^{-4} - 10^{-5}$, which are typically within reach of the LHC detectability expectations \cite{Carvalho:2007yiHan:2008xb}. For an integrated luminosity of 10 fb$^{-1}$ this branching ratio can be explored down to $3.1\times 10^{-4}$, for an integrated luminosity of 100 fb$^{-1}$ this figure is pushed to $6.1\times 10^{-5}$. Together with $t\to c Z$, the decays $t\to u Z$ are also of potential interest. Nevertheless, the corresponding branching ratio is much smaller -- typically $\mathcal O(10^{-6})$ --, as could be anticipated from the bounds on $|\Uu{14}|$ being tighter than on $|\Uu{24}|$. Notice once again that these branching ratios of $t\to q Z$ decays are not unavoidably obtained with such a size: they would automatically drop down for mixing matrices much closer to the $3\times 3$ unitary case. It is however true that once we focus on the possibility of obtaining significant phases in \BBsmix\ mixing, then sizable values of $\text{Br}(t\to Zc)$ follow.

By the same token, $|\V{tb}|$ is sizeably different from unity, affecting single top production at hadron machines and other observables.
\\

Once we have imposed the different constraints explained above, there are other rare decays that can be slightly enhanced. This is the case of $K_L\to \pi^0\nu\bar\nu$ and $B_{s,d}\to\mu^-\mu^+$. The predictions are always below the actual experimental bounds\footnote{A more detailed analysis will be presented in a future paper.}.

At this stage, it is worth mentioning that recently an up-type isosinglet quark extension of the SM has been considered in the literature \cite{Alwall:2006bxPicek:2008dd,Vysotsky:2006fxKopnin:2008ca}, where it is assumed that the heavy quark $T$ only mixes with the top quark. One can readily see that effects such as sizable $\chi$, $t\to cZ$ decay rates or \DDmix\ mixing contributions, proportional to $|\Uu{24}\Uu{34}|^2$ and $|\Uu{14}\Uu{24}|^2$ respectively, are absent.

\section{Summary and Conclusions}\label{SEC:CONC}

We have investigated the question of how to accommodate the large value of $\chi_{eff}$ (of order $\lambda$) recently observed in the flavour-tagged determination of mixing induced CP violation in $B_s\to J/\psi\phi$ decays. If confirmed, this value would signal the presence of New Physics, since in the SM, one necessarily has $\chi=\chi_{eff}=\mathcal O(\lambda^2)$. We have pointed out that in models with small violations of $3\times 3$ unitarity, one can have $\chi_{eff}$ of order $\lambda$ provided the new quark $T$ mixes significantly with the top and also with the charm quark. This enhancement is in part due to an important correction in the exact relation between $\chi$ and other rephasing invariant quantities. 
We have carefully examined the implications of the model for various important observables which severely constrain any extension of the SM. We have found that in order to have $\chi$ of order $\lambda$, while conforming to the above constraints, the mass of the $T$ quark cannot exceed around $500$ GeV. It is clear that other New Physics may also lead to $\chi_{eff}=\mathcal O(\lambda)$. For example, in some extensions of the SM, including the supersymmetric ones \cite{Branco:1994ebBranco:1995cj,Masiero:2001nbMasiero:2003fy,Altmannshofer:2007cs}, there are New Physics contributions to \BBsmix\ mixing which may in principle lead to $\chi_{eff}>\lambda^2$ even if $\chi$ is of order $\lambda^2$. Recently there has been a great deal of interest in investigating flavour implications, including a possible enhancement of $\chi$ in theories beyond the SM, such as Little Higgs models \cite{ArkaniHamed:2001nc,littlehiggs1,Blanke:2006eb}, four generations \cite{fourgenerations} and supersymmetric extensions \cite{Dutta:2008xgHisano:2008df}.
The distinctive feature of the present model is the fact that a value of $\chi=\mathcal O(\lambda)$ necessarily implies rare top decays $t\to cZ$ at a rate observable at the LHC. We have also examined the question of \DDmix\ mixing. In the present framework, there are new contributions to this mixing, arising at tree level from Z-mediated flavour changing neutral currents. We have pointed out that even taking into account the limits on deviations of unitarity arising from normalization of the first row of $\CKM$, one can have values of $\Uu{14}$ leading to contributions to \DDmix\ mixing which can account for the present experimental value, without having to invoke long distance contributions. It was emphasized that, contrary to the case of $t\to cZ$ decays, this is not a mandatory feature of the present model, since the new contributions to \DDmix\ mixing can be made arbitrarily small by choosing $\Uu{14}$ sufficiently close to zero.\\
In conclusion, the presence of small unitarity violations may have important effects on the size of $\chi$ and other measurable quantities, which will be probed with higher precision at LHCb and future superB factories.

\section*{Acknowledgements}\label{SEC:Ack}
The authors acknowledge J. Portol\'es for useful conversations. This research has been supported by European FEDER, Spanish MEC under grant FPA 2005-01678, PORT2007-03 and SAB2006-0044, as well as by \emph{Funda\c{c}\~{a}o para a Ci\^{e}ncia e a Tecnologia} (FCT, Portugal) through the projects\\ {\small PDCT/FP/63912/2005}, {\small PDCT/FP/63914/2005}, {\small POCI/FP/81919/2007} and {\small CFTP-FCT UNIT 777} which are partially funded through POCTI (FEDER), and by the Marie Curie RTN {\small MRTN-CT-2006-035505}. 

\appendix
\section{Numerical input}\label{AP:Num}
Table \ref{TAB:EXP} collects the experimental values of the observables that have been used \cite{hfag,pdg,Towner:2007npEronen:2007qc,Antonelli:2008jg,Aba:2008fj,Aubert:2008bd,Aubert:2007wfStaric:2007dtAbe:2007rd,Carvalho:2007yiHan:2008xb,Eusebi:2008ju,Iwasaki:2005syAubert:2004it}.

\begin{table}[h]
\begin{center}
\begin{tabular}{|c|c||c|c|}
\hline
  Observable & Exp. Value & Observable & Exp. Value \\ \hline\hline
$|\V{ud}|$ & $0.97408\pm 0.00026$ & $|\V{us}|$ & $0.2253\pm 0.0009$\\\hline
$|\V{cd}|$ & $0.230\pm 0.011$ & $|\V{cs}|$ & $0.957\pm 0.095$\\\hline
$|\V{ub}|$ & $0.00431\pm 0.00030$ & $|\V{cb}|$ & $0.0416\pm 0.0006$\\\hline
$\gamma$ & $(76\pm 23)^\circ$ \\\hline
$\AJPsi$ & $0.675\pm 0.026$ &  $\AJPsiP$ & $0.540\pm 0.225$\\\hline
$\DMBd$($\times$ ps) & $0.507\pm 0.005$ & $\DMBs$ ($\times$ ps)& $17.77\pm 0.12$\\\hline
$x_D$ & $0.0097\pm 0.0029$ & $\Delta T$ &  $0.13\pm 0.10$\\\hline
$\epsilon_K$($\times 10^3$) & $2.232\pm 0.007$ & $\epsilon^\prime/\epsilon_K$($\times 10^3$) & $1.67\pm 0.16$\\\hline
Br$(\Kpnn)$ &  $(1.5\begin{smallmatrix}+1.3\\ -0.9\end{smallmatrix})\times 10^{-10}$ & Br$(K_L\to\mu\bar\mu)_{SD}$ & $<2.5\times 10^{-9}$ \\\hline
Br$(B\to X_s \ell^+\ell^-)$ & $(1.60\pm 0.51)\times 10^{-6}$ \\\hline
Br($t\to cZ$) & $<4\times 10^{-2}$ & Br($t\to uZ$) & $<4\times 10^{-2}$ \\\hline
$\DGBs$ ($\times$ ps)& $0.19\pm 0.07$& $\DGBsCP$ ($\times$ ps)& $0.15\pm 0.11$ \\\hline
$\DGBdGBd$ & $0.009\pm 0.037$\\\hline
$\Asld$ & $-0.003\pm 0.0078$ &$\Asl$ & $-0.0028\pm 0.0016$\\\hline
\end{tabular}
\caption{Experimental values of observables.}\label{TAB:EXP}
\end{center}
\end{table}

\section{$\boldsymbol{\AJPsiP}$ and the contribution of $\boldsymbol{\chi}$}\label{AP:signCHI}

The mixing induced CP violating asymmetry in the CP even part of the final state $B_s\to J/\Psi\Phi$ is $\AJPsiP=\sin 2\chi_{eff}$, where $\chi_{eff}$ is defined in the \BBsmix\ mixing matrix. Factorizing the modulus of the SM expression $|[M^s_{12}]_{SM}|$, we have
\begin{eqnarray}
 \frac{M_{12}^s}{|[M^s_{12}]_{SM}|}&=&r_s^2 e^{-i2\chi_{eff}} \label{EQ:M12Bs}\\
&=&e^{-i2\chi}\left\{1+2\,\frac{S(x_t,x_T)}{S(x_t)}\frac{\Vc{Ts}\V{Tb}}{\Vc{ts}\V{tb}}+\frac{S(x_T)}{S(x_t)}\left(\frac{\Vc{Ts}\V{Tb}}{\Vc{ts}\V{tb}}\right)^2\right\}~,\nonumber\\
&=&e^{-i2\chi}~r_s^2~e^{-i2\varphi}~.\nonumber 
\end{eqnarray}

It is clear that $\AJPsiP$ has two contributions: one from $\chi$ and the second coming from the new heavy top quark $T$ running inside the box diagram, namely $\varphi$. The Inami-Lim \cite{Inami:1980fz} functions $S$ verify $S(x_t,x_t)=S(x_t)$. In order to understand the behaviour of these two contributions let us stress the following facts:
\begin{itemize}
 \item In reference \cite{AguilarSaavedra:2004mt} we have shown the existence of a \emph{screening}: in the limit $m_T\to m_t$, and independently of the value of $\chi$, we have $\chi+\varphi=\mathcal O(\lambda^2)$, as in the SM. More precisely,
\begin{equation}
 \lim_{m_T\to m_t} \chi+\varphi=\arg\left(1+\frac{\Vc{ub}\V{us}}{\Vc{cb}\V{cs}}\right)\leq \arcsin\left(\frac{|\Vc{ub}\V{us}|}{|\Vc{cb}\V{cs}|}\right)\sim\mathcal O(\lambda^2)~.\label{EQ:Screening}
\end{equation}
So in the limit $m_T\to m_t$, $\varphi$ is either order $-\chi$ -- for $\chi\sim\lambda$ --, or very small -- for $\chi\sim\lambda^2$ --.
 \item The phase $\varphi$ is dominated by the imaginary part of $-\frac{S(x_t,x_T)}{S(x_t)}\frac{\Vc{Ts}\V{Tb}}{\Vc{ts}\V{tb}}$. Note that the ratio of CKM matrix elements is at most of order $\lambda$.
 \item The $x_T$ dependence of the Inami-Lim function $S(x_t,x_T)$ destroys the cancellation in \eq{EQ:Screening} as soon as we move away from $m_T=m_t$. For example, $S(x_t,x_T)\simeq 1.5 S(x_t)$ for $m_T=300$ GeV, and growing with $m_T$. 
\end{itemize}
So we conclude that to have $\chi_{eff}$ of order $\lambda$, we have to pick $\chi$ of order $\lambda$ and, as soon as the mass of the new quark $T$ grows, we will get $\chi_{eff}$ of the same order and opposite size. That is the reason why in all our examples, to have $\chi_{eff}$ of order $\lambda$ and positive, we need $\chi$ of order $\lambda$ but negative. So it turns out that to have a large positive $\AJPsiP$ one has to start with a large negative $\chi$.

\newpage

\end{document}